\documentclass[twocolumn,showpacs,preprintnumbers,aps,amsmath,amssymb]{revtex4}

\usepackage{graphics,graphicx}
\usepackage{dcolumn}
\usepackage{bm}
\begin{document}

\title{Observation of condensed phases of quasi-planar core-softened colloids}
\author{N. Osterman$^1$, D. Babi\v c$^1$, I. Poberaj$^1$, J. Dobnikar$^{2}$, and P. Ziherl$^{1,2}$}
\affiliation{$^1$Department of Physics, University of Ljubljana, Jadranska 19, SI-1000 Ljubljana, 
Slovenia}
\affiliation{$^2$Jo\v zef Stefan Institute, Jamova 39, SI-1000 Ljubljana, Slovenia}
%\affiliation{$^*$ Electronic adress: jure.dobnikar@ijs.si}
\date{\today}

\begin{abstract}
We experimentally study the condensed phases of repelling core-softened spheres in two dimensions. The 
dipolar pair repulsion between superparamagnetic spheres trapped in a thin cell is induced by a 
transverse magnetic field and softened by suitably adjusting the cell thickness. We scan a broad density 
range and we materialize a large part of the theoretically predicted phases in systems of core-softened 
particles, including expanded and close-packed hexagonal, square, chain-like, stripe/labyrinthine, and 
honeycomb phase. Further insight into their structure is provided by Monte Carlo simulations.
\end{abstract}

\maketitle

One of the ageless challenges of material science is to engineer structure formation. Recently, of 
particular interest are the various crystalline phases of colloidal particles~\cite{Xia00} which may 
be used for self-assembled nanomaterials~\cite{Fendler96}, photonic crystals~\cite{Busch98}, macroporous 
polymer membranes~\cite{Jiang99}, biochemical applications~\cite{Kawaguchi00}, etc. In this context, it 
is imperative to understand the link between the interparticle potential and the phase behavior of the 
ensemble. Experimentally, this link can be studied readily using table-top equipment which provides the 
full microscopic real-time structural information, wherefrom the subtle relations between the microscopic 
and the macroscopic properties of the system can be reconstructed. Complementary insight can be obtained 
by inverse methodology to identify the optimal pair potential that produces a given target 
structure~\cite{Rechstman05}. 

A basic limitation of these efforts is the spherical shape typical for many colloids and the ensuing 
isotropy of the interparticle potential. This is the main reason why the most common ordered structures 
are close-packed, i.e., the hexagonal and the face-centered cubic 
lattice~\cite{Kremer86,Mau99} in two and three dimensions, respectively. Conceivably, a richer phase 
diagram can be induced by anisotropic interactions due to, e.g., surface treatment of 
particles~\cite{Nelson02}, external fields~\cite{Yethiraj03} or a liquid-crystalline solvent~\cite{Stark01}. 
Another route to the more open lattices are isotropic pair interactions with a radial profile characterized 
by two length scales, such as the combination of hard-core and penetrable-sphere repulsion~\cite{Malescio03}. 
If the shoulder/core diameter ratio exceeds about 2, this pair potential stabilizes a range of mesophases 
intervening between the fluid and the close-packed crystal. In two dimensions, the theoretical phase sequence 
includes loose- and close-packed hexagonal lattice; monomer, dimer, and trimer fluids; stripe and labyrinthine 
phases; honeycomb lattice, etc.~\cite{Malescio03,Norizoe05,Jagla99}. Very similar behavior has been predicted 
numerically in paramagnetic particles confined to a plane and interacting with a dipolar repulsion induced by 
a transverse magnetic field and softened by a Lennard-Jones interaction~\cite{Camp03}. For large shoulder/core 
diameter ratios, the set of mesophases reduces to micellar, lamellar, and inverted micellar structure~\cite{Glaser07}.

The theoretical insight into the mesophases formed by particles with a core-softened isotropic repulsive 
pair potential in two dimensions is reasonably comprehensive but the physical evidence of their stability is 
still lacking. In this Letter, we study the phase sequence of such a system experimentally using superparamagnetic 
spheres, tailoring the profile of interaction with external magnetic field and spatial constraints. We discover 
a host of the predicted structures, and we corroborate the observations with numerical simulations. 

We used a thin wedge-shaped cell formed by nearly parallel glass plates and filled by a water solution of 
$1.05~{\rm \mu m}$ superparamagnetic spheres (Dynabeads, MyOne Carboxy). A small (5~mg/ml) amount of SDS 
was added to suppress the van der Waals interactions and prevent sticking of the colloids. Like in related 
experiments~\cite{Bubeck98,Zahn99,Wen00,Eisenmann04,Erbe07}, the interparticle repulsion was induced by external 
magnetic field. The experimental setup is comprised of laser tweezers system built around an inverted 
optical microscope (Zeiss Axiovert 200M) and a CW diode pumped Nd:YAG laser (Coherent, Compass 2500MN), 
and equipped with three orthogonal pairs of Helmholtz coils driven by a computer-controlled current source. 
Multiple laser traps used to manipulate the spheres are created by time-sharing using acousto-optic deflectors 
(IntraAction, DTD-274HA6) driven by beam-steering controller (Aresis, Tweez). The cell is illuminated with 
a halogen lamp and the micrographs are recorded in the bright field with a fast CMOS camera (PixeLINK, 
PL-A741). Image analysis is performed offline with custom-made particle tracking software. 

\begin{figure}[ht]
\includegraphics[width=0.42\textwidth]{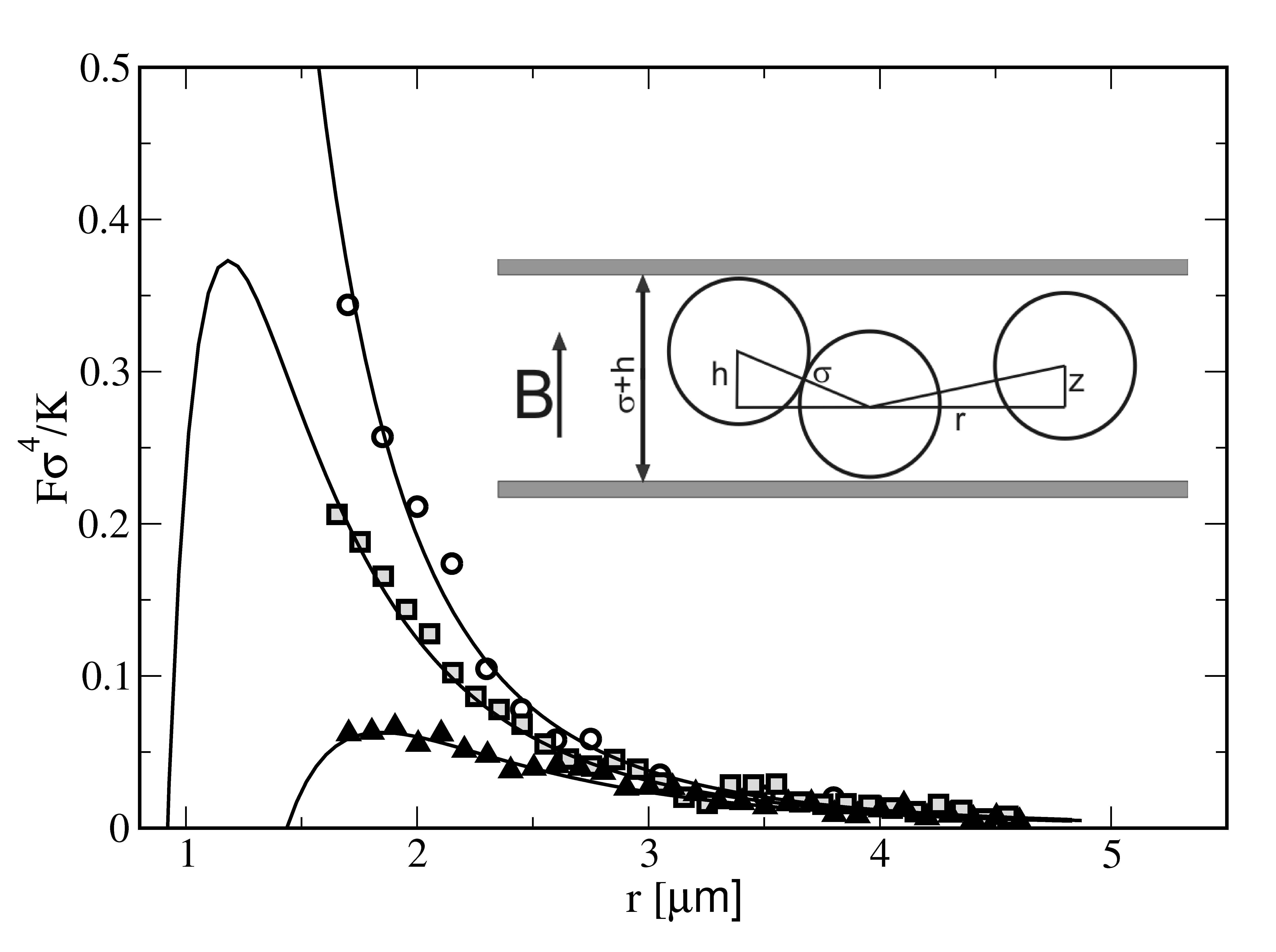}
\caption{Measured variability of the in-plane pair force profiles: $r^{-4}$ repulsion (thin cell, $h\ll h_m$; 
circles), softened repulsion ($h\approx h_m$; squares), and oversoftened interaction (thick cell, $h>h_m$; 
triangles) with the attractive part at small separations. For $h=h_m$, the in-plane hard-core diameter of the 
spheres is $\approx0.89~\mu{\rm m}$. The small-separation part of the force profile is not accessible with our 
method as the laser would act on both spheres, rendering the measurement impossible. In our experiment, $B$ 
is typically around  10~mT and the induced pair forces are of the order of 0.1~pN at separations comparable 
to sphere diameter. --- Solid lines are theoretical fits for $h=0$, $h=0.46~\mu{\rm m}$ and $h=0.72~\mu{\rm m}$, 
respectively. Inset: Experimental geometry in cross-section.}
\label{Pot}
\end{figure}

The system's key feature which softens the repulsion between the induced magnetic dipoles of the spheres is 
the cell thickness: If it is somewhat larger than the sphere diameter, the centers of two near-by spheres are 
not restricted to a plane perpendicular to the magnetic field. This makes their interaction at small separations 
less repulsive (or even attractive) compared to spheres lying in the plane. The pair potential is
\begin{equation}
U(r,z)= K\frac{r^2-2z^2}{\left(r^2+z^2\right)^{5/2}},
\label{Ur}
\end{equation}
where $K=\pi\sigma^6\chi^2 B^2/144\mu_0$ is the interaction constant which depends on the magnetic induction 
$B$ and the magnetic susceptibility $\chi$ of the particles; $r$ and $z$ are the in-plane and the vertical 
separations of the spheres, respectively, and $\sigma$ is their diameter (Fig.~\ref{Pot}). It is convenient 
to write the cell thickness as $\sigma+h$: Thus $h$ measures the deviation from a truly planar system. In the 
thin part of the cell where $h\to0$, $U$ reduces to $K/r^3$, whereas for $h>0$ the repulsion is reduced 
at small distances by the relative vertical shift of the spheres. The in-plane interparticle 
force is $F_r=-\partial U/\partial r=-3Kr\left(4z^2-r^2\right)/\left(r^2+z^2\right)^{7/2}.$ For sphe\-res in 
contact in a cell of thickness $\sigma+h<2\sigma$, $F_r=-3K\sqrt{\sigma^2-h^2}\left(5h^2-\sigma^2\right)/
\sigma^7$ which is attractive for $h>h_m=\sigma/\sqrt{5}\approx0.447\sigma$. The softening of the repulsion 
is most pronounced at cell thicknesses somewhat smaller than $\sigma+h_m$.

To verify that the in-plane pair interaction between the spheres is indeed described by the above law, we 
measured it using two isolated spheres, one attached to the lower glass plate and the other one (probe) 
confined laterally by a weak optical trap and pushed against the upper plate by the light pressure so that 
their vertical separation was fixed to $h$. The probe was repeatedly slowly dragged back and forth from the 
fixed sphere, and its trajectories were recorded and averaged to accurately determine the displacement of the 
probe particle from the trap center as a function of interparticle distance. To calculate the magnetic force 
on the probe from the displacement profile, the stiffness of the laser potential was determined by analyzing 
the Brownian motion of the probe sphere with the magnetic field turned off. The procedure was repeated at a 
range of cell thicknesses and the three representative force profiles shown in Fig.~\ref{Pot} demonstrate 
that the system behaves as expected. 

Having confirmed that the pair potential is given by Eq.~(\ref{Ur}), we studied the phase behavior of the system 
across a broad density range. We filled the cell with a dense suspension of spheres and we first located the 
measuring site with cell thickness of about $\sigma+h_m$. As the cell thickness cannot be measured directly, 
we used the following procedure. In absence of the magnetic field, we herded the spheres at a given location 
into contact such that they formed clusters. We then turned on the magnetic field which caused the clusters to 
disintegrate into smaller chunks. In the thick part of the cell where the potential was attractive at small 
separations the cluster disintegration was partial, whereas in the thin part with purely repulsive interactions 
the clusters disintegrated completely. By scanning the region we identified the site with the desired cell 
thickness.

At this site, we increased the sphere density $n$ in the part of the cell covered by camera's field of view 
using the laser tweezers operating at high power to locally heat the suspension. The heating induced a
hydrodynamic flow which dragged a large number of spheres (typically over $10^4$) towards the trap, thereby
increasing the local sphere density almost to close-packing. Then the laser was turned off which stopped 
the flow, and the magnetic field was turned on. The high-density region underwent an adiabatic expansion 
so the system could be studied at decreasingly lower colloidal densities. A sizable change of the density 
took place on a time scale of about 10~s. This is much longer than the typical diffusion time of particles 
$<10~{\rm ms}$~\cite{timescale} so that the expansion was slow enough to ensure quasi-equilibrium at all times, 
and the hydrodynamic interactions due to expansion $\sim10^{-3}k_BT$ are negligible compared to the magnetic 
repulsion. 

\begin{widetext}

\begin{figure}[t]
\includegraphics[width=0.95\textwidth]{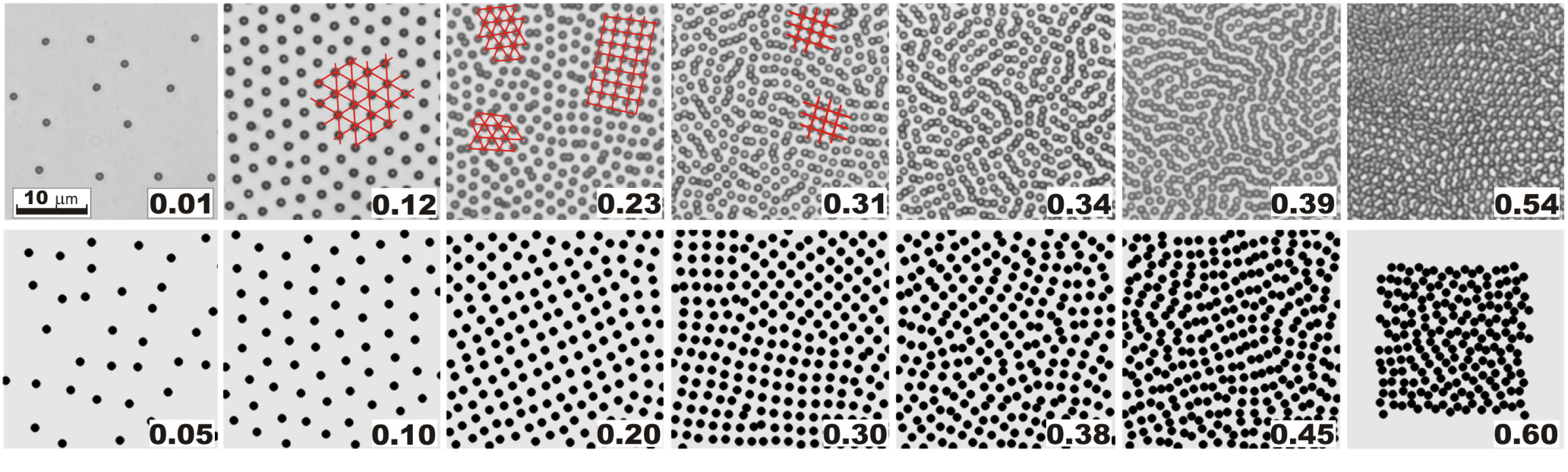}
\caption{Micrographs of the representative mesophases induced by varying the density (top row; $B=12.5~{\rm mT}$ 
which gives $k\approx400$) and the simulation snapshots (bottom row) labelled by the corresponding 2D volume 
fractions of colloids. Due to optical image distortion, the colloids on the micrographs appear slightly larger 
than their true size; in some of the micrographs, patches of the underlying lattices are emphasized to guide the eye. 
--- The reduced interaction strength of $k=375$ was used in simulations at all but the largest two volume fractions 
where $k$ was set to $125$ to improve convergence; the number of particles $N=1000$ except at $\eta=0.60$ where $N=200$.}
\label{Snap}

\end{figure}
\end{widetext}

Micrographs of the most interesting phases observed at various 2D volume fractions $\eta=\pi\sigma^2 n/4$ shown 
in Fig.~\ref{Snap} include i) fluid phase ($\eta=0.01$); ii) expanded hexagonal lattice ($\eta=0.12$); iii) 
coexisting expanded hexagonal and square lattice ($\eta=0.23$); iv) coexisting expanded square lattice and chain 
phase ($\eta=0.31$); v) chain phase with locally aligned finite-length strings ($\eta=0.34$); vi)~stripe/labyrinthine 
structure formed predominantly by a single cluster of interlaced strings of touching spheres ($\eta=0.39$); and vii) 
coexisting honeycomb and dense square lattice at very high volume fraction ($\eta=0.54$). The observed mesophases 
are remarkably close to those found in the numerical simulations reported in Ref.~\cite{Camp03} although the pair 
interaction is not exactly the same; a much more idealized hard-core/soft-shoulder interaction also gives a similar 
phase sequence~\cite{Jagla99,Malescio03}. This suggests that the mechanisms at work as well as the structures they 
produce are rather robust. 

Nonetheless, our system is not perfectly two-dimensional, and the vertical position of a sphere is a restricted 
variable which does affect the symmetry of the in-plane equilibrium configuration. To understand the phase sequence 
in more detail, we performed 3D Monte Carlo simulations of up to $N=1000$ spheres with the dipole-dipole pair 
repulsion in a cell of thickness $\sigma+h_m$. We varied the volume fraction and we focused on the low temperature 
limit corresponding to large reduced interaction strengths  $k=K/k_BT\sigma^3$; in the experiment, $k\approx400$. 
We used periodic boundary conditions in several simulation box geometries; after reaching equilibrium (typically 
in a few million steps) $50\times10^6$ averaging steps were performed to evaluate the energy per particle, the 
radial distribution function, and the static structure factor. The selected simulation snapshots in Fig.~\ref{Snap} 
are chosen to best reproduce the phase sequence seen in the micrographs. The agreement is very good, which also 
demonstrates that any effect of polydispersity of the superparamagnetic spheres used is unimportant.

The insight provided by the simulations goes beyond the experimental micrographs: It includes the vertical positions 
of the spheres, which are uncorrelated and evenly distributed across the available range as long as the volume 
fraction does not exceed about 0.05. However, for $\eta\gtrsim0.1$ the distribution becomes bimodal with pronounced 
peaks corresponding to spheres touching the bottom and the top plate, and the system resembles an off-lattice 
two-state spin ensemble with dominant nearest-neighbor antiferromagnetic interactions. The nearest-neighbor 
interactions are frustrated by the hexagonal lattice but the stripe, honeycomb, and square lattice with 2, 3, and 4 
regularly arranged nearest neighbors, respectively, are compatible with alternating up-down positions of spheres. 
We calculated the energies of the stripe, square, and hexagonal phase by fixing their in-plane structure~\cite{stripes}. 
The ground states of the non-frustrated phases, such as the checkerboard structure of the square lattice, are ordered 
and unique, and their energies were calculated using a lattice sum, whereas the energy of the hexagonal lattice was 
evaluated by annealing the vertical positions numerically in systems of up to $N=5000$ spheres. 

\begin{figure}[t]
\includegraphics[width=0.42 \textwidth]{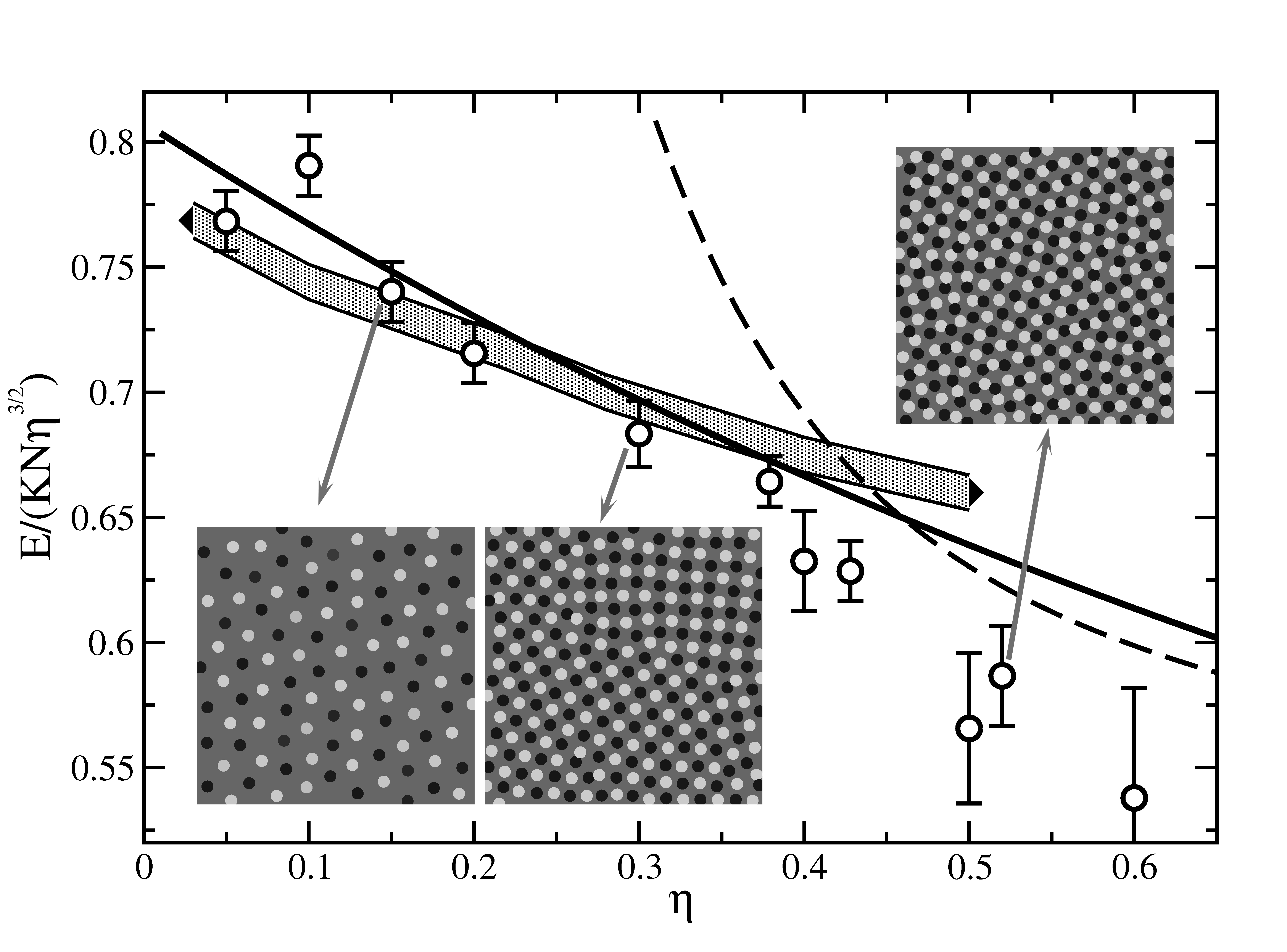}
\caption{Energy per particle of the ground states of hexagonal (broad shaded line; the line thickness represents 
the numerical inaccuracy estimated by varying both the number of particles and the coupling constant), square (solid 
line), and stripe phase (dashed line) compared to the energies obtained by the full Monte Carlo simulation (open 
circles with error bars). Also shown are the snapshots for $\eta=0.10$ (hexagonal lattice), $\eta=0.30$ (square 
lattice), and $\eta=0.52$ (labyrinthine structure). The spheres' vertical positions are encoded by shades of gray 
(light gray: $z=0$, black: $z=h$). The checkerboard pattern of the square lattice is clearly visible, and so is the 
alternating up-down intra-stripe order of the labyrinthine structure. In the hexagonal lattice, the vertical positions 
of the spheres are disordered.}
\label{PhD}
\end{figure}

The results are shown in Fig.~\ref{PhD}: At small $\eta$, the hexagonal phase has the lowest energy per particle 
among the three candidate phases, the square phase is stable at intermediate volume fractions, and at largest volume 
fractions considered the stripes are energetically most favorable. In Fig.~\ref{PhD}, we also plotted the energies of 
the equilibrium states obtained by the full Monte Carlo analysis that produced the snapshots shown in Fig.~\ref{Snap}. 
Up to $\eta\approx0.4$, the agreement of the energies calculated using lattice sums and simulations, respectively, 
is quite good. In this regime, the simulations as well as the experiments usually produce coexistence of the 
hexagonal and the square lattice rather than a pure lattice (Fig.~\ref{Snap}). This can be understood based on the
lattice sum results which predict a very similar dependence of the energies of the two lattices on $\eta$, and thus 
the coexistence regime obtained by Maxwell construction should extend over a broad density range. --- At volume 
fractions beyond $\approx0.4$, the difference between the energies of the model structures and those obtained by 
simulations slightly exceeds the error bar of the latter suggesting that the stripe phase with perfectly parallel 
straight stripes is probably not the ground state and many turns and junctions may be an equilibrium feature of the 
system. 

In conclusion, we report an experimental study where a core-softened repulsive interaction was induced in 
micrometer-size colloidal particles using the external magnetic field and fine-tuned by controlling the spatial 
constraints. Depending on the density, the system forms several self-assembled mesophases --- the square, hexagonal, 
and honeycomb lattices as well as the labyrinthine structure ---, thereby experimentally validating the 
theoretical predictions pertaining to similar pair interactions~\cite{Jagla99,Malescio03,Camp03}. The observed 
two-dimensional structures could be used for the fabrication of templates to promote the growth of colloidal 
crystals~\cite{vanBlaaderen97,Lin00}.

We thank H. H. von Gr\"unberg for stimulating discussions and B. Kav\v ci\v c for technical assistance. 
This work was supported by the European Commission (J. D., Grant MERG-031089 and 
by the Slovenian Research Agency (D. B. and I. P., Grant J1-6502; J. D. and P. Z., Grant P1-0055).


\begin{thebibliography}{99}
\bibitem{timescale} In equilibrium, a thermally excited particle in the potential well formed by its neighbors
fluctuates around its mean position by about $d\sim[k_BT\sigma^5f(\eta)/384K]^{1/2}$ where $f(\eta)$ is a
numerical factor which depends on the volume fraction and equals 1 at close-packing and $\sim1000$ at 
$\eta\approx0.1$. The typical time needed to traverse this distance is $\tau\sim d^2/D$, where 
$D=k_BT/3\pi\nu\sigma$ is the diffusion constant; $\nu$ is the viscosity. In our case, $\tau\sim10~\mu{\rm s}$ 
at $\eta=0.5$ and $\tau\sim10~{\rm ms}$ at $\eta=0.1$.
\bibitem{stripes} We analyzed all stripe structures of width~1 by varying the intra-stripe separation, the 
stripe spacing, and the staggering of neighboring stripes. At any volume fraction, the minimal-energy 
configuration consists of close-packed stripes, and the neighboring stripes are out of registry by 1 sphere 
diameter such that the stripe structure resembles a square lattice compressed along one lattice vector and 
stretched along the other one as much as possible.
\end{thebibliography}
\end{document}